\documentclass[10pt, letterpaper]{article}
\usepackage{graphicx}
\usepackage{dcolumn}
\usepackage{bm}
\usepackage[margin=2.5cm]{geometry}

\usepackage{xcolor}
\usepackage[normalem]{ulem}
\usepackage{amsfonts}
\usepackage{mathtools}

\newcommand{\ud}{\mathrm{d}}
\newcommand{\ue}{\mathrm{e}}
\newcommand{\ui}{\mathrm{i}}

\begin{document}

\title{Are inertial vacua equivalent in Lorentz-violating theories?\\
 Does it matter?}

\author{Bruno Arderucio Costa$^1$, Yuri Bonder$^2$, Benito A. Ju\'arez-Aubry$^3$\\
Instituto de Ciencias Nucleares, Universidad Nacional Aut\'onoma de M\'exico\\
Apartado Postal 70-543, Cd.~Mx. 04510 Mexico\\
\small{$^1$bruno.arderucio@correo.nucleares.unam.mx $^2$bonder@nucleares.unam.mx $^3$benito.juarez@correo.nucleares.unam.mx}}
 
\date{\today}
\maketitle

\begin{abstract}
Several approaches to quantum gravity suggest violations of Lorentz symmetry as low-energy signatures. This article uses a concrete Lorentz-violating quantum field theory to study different inertial vacua. We show that they are unitarily inequivalent and that the vacuum in one inertial frame appears, in a different inertial frame, to be populated with particles of arbitrarily high momenta. At first sight, this poses a critical challenge to the physical validity of Lorentz-violating theories, since we do not witness vacuum excitations by changing inertial frames. Nevertheless, we demonstrate that inertial Unruh-De Witt detectors are insensitive to these effects. We also discuss the Hadamard condition for this Lorentz-violating theory.
\end{abstract}

\section{Introduction}

One of the main challenges in theoretical physics is to find a theory that consistently incorporates quantum and gravitational physics. The attempts to construct such a theory can be grouped into two categories. In one, theories that incorporate some quantum aspects into gravitation, or vice versa, are proposed from first principles. This strategy includes theories like string theory, loop quantum gravity, and other remarkable ideas (for a review of several quantum gravity approaches, see Ref.~\cite{Oriti}). In the other strategy, physicists consider general relativity as an effective low-energy description of a more fundamental theory, and they look for experimental evidence of quantum gravity. Since we have not detected them yet, these signatures must manifest themselves, in the accessible regimes, as small deviations from the predictions of the current paradigm.

To empirically search for discrepancies with general relativity, it is useful to have a parameterisation. In turn, this parameterisation can be generated by assuming violations of some fundamental principle. One of the most important principles of current physics is invariance under Lorentz transformations. This principle states that all inertial observers are equivalent, thus preventing the existence of preferred spacetime directions. Importantly, several promising quantum gravity candidates argue that Lorentz invariance is broken, in one way or another, close to the Planck scale~\cite{KosteleckySamuel, GambiniPullin}. Regardless of the fundamental mechanism to break Lorentz invariance, it is relevant to study its empirical and conceptual implications. This has been done following several strategies. For example, the Einstein-Aether programme~\cite{Mattingly} breaks Lorentz invariance by introducing a timelike vector field that can be associated with a preferred reference frame. In this approach, such a field is dynamical, leading to a spontaneous breakdown of Lorentz invariance. It is also possible, as we do here, to consider non-dynamical tensor fields to introduce explicit violations of this principle~\cite{Cristobal1, Cristobal2}. 

There exists a model-independent framework to study violations of Lorentz invariance, from hereon referred to as Lorentz violation. The idea is to create a parameterisation in the effective field theory framework~\cite{Rob}; the result is known as the Standard Model Extension (SME)~\cite{SME1, SME2, Kostelecky2004}. As such, the SME action contains that of conventional physics (i.e., the Standard Model and general relativity), plus all gauge-preserving corrections that can be written with the fields of this conventional part, but which are not necessarily Lorentz invariant. The Lorentz-violating terms are weighed by non-dynamical tensor fields, dubbed SME coefficients.

There is vast literature exploring a variety of Lorentz-violating effects. For example, there are several hundred different experiments that have been used to set bounds on the SME coefficients (for reviews on these bounds see Refs.~\cite{Mattingly2005-au, Liberati_2013, DataTables}). The most frequently studied effect associated with Lorentz violation is the appearance of modified dispersion relations. Also, people have analysed theoretical consequences related to Lorentz violation~\cite{PhysRevD.63.065008}, particularly when considering dynamical spacetime geometries and/or spontaneous Lorentz violation~\cite{SMEtheo1, SMEtheo2, SMEtheo3, SMEtheo4, SMEtheo5, SMEtheo6, SMEtheo7, SMEtheo8}. 

Unfortunately, the quantisation of Lorentz violating theories has not been studied with the same depth. One notable exception is Ref.~\cite{MauroRalfRob}, where radiative corrections and the running of the SME coefficients are studied. Yet, in this reference, the symmetries of the vacuum, which are the main topic of this paper, are not analysed. We also refer the reader to Ref.~\cite{Husain16}, where a detector's response to a quantum field with an unconventional dispersion relation, where the modifications are suppressed by a large mass, is analysed.

In this paper, we quantise a free Lorentz-violating scalar coupled to a dimensionless SME-inspired coefficient, which is the simplest Lorentz-violating theory one can consider (Sec.~\ref{secModel}). Of course, we could study a more realistic theory, namely, one that is actually within the (noninteracting sector of the) SME, but we do not foresee any advantage in doing so as the goal of the paper can be conceptually achieved in this simpler scenario.

We then find the inertial vacuum states of this theory (Sec.~\ref{Sec:Vacua}). We find that, due to the presence of a non-dynamical tensor field, the vacuum state in a particular Lorentz frame is not unitarily connected with a vacuum associated with another inertial frame. This result is in stark contrast with the Lorentz-invariant counterpart, where the Minkowski vacuum is Poincar\'e invariant. Moreover, the action of the Poincar\'e group is implemented, at the quantum level, by unitary transformations, ensuring that the same quantum theory is described in different inertial frames.

What we find would raise serious questions on the phenomenological viability of Lorentz-violating theories, even if they are merely seen as low-energy effective theories. To investigate the consequences of our findings, we study whether different inertial observers can detect such effects (Sec.~\ref{Sec:Detector}). To probe the particle content of an inertial vacuum in different inertial frames, we equip our inertial observers with Unruh-De Witt detectors. We find that, despite the initial concerns, inertial detectors remain silent when travelling through the vacuum as defined by another inertial frame. Following Unruh's motto that `a particle is what a particle detector detects,' in this simple Lorentz-violating model, any inertial observer experiences the vacuum associated with one fixed inertial frame as devoid of particles.

Eventually, it may be interesting to extend the above results to perturbative interacting theory. Thus, we complement the aforementioned results with a discussion on the Hadamard property in this Lorentz-violating theory (Sec.~\ref{sec:Hadamard}). This condition, as it turns out, does not strictly hold and needs to be adapted. Nevertheless, we see no obstruction in using this `modified' Hadamard property to construct interacting observables (including time-ordered products of non-linear observables) and explore these questions in perturbative interacting theory. We end this paper with some concluding remarks (Sec.~\ref{sec:conc}).

Throughout the paper, Latin indices from the beginning of the alphabet represent abstract indices~\cite{WaldGR}, which stand for open slots in the associated tensors. Greek indexes represent the components of the corresponding tensors in a given coordinate basis. For simplicity, the spacetime under consideration is the flat $4$-dimensional spacetime (without torsion or non-metricity), whose metric (resp. inverse metric) is denoted by $\eta_{ab}$ (resp. $\eta^{ab}$), and it has signature $(-\ +\ +\ +)$. The covariant derivative is denoted by $\nabla_a$ and complex conjugation is denoted by a bar over the symbol.

\section{The model}\label{secModel}

For our purposes, it is enough to study the simplest possible setting, which is given by quantum scalar field $\phi$. Of course, this field is not directly part of the SME. Still, this model contains valuable lessons that can be generalised for realistic matter fields.

Inspired by the notation in the spinorial sector of the SME, we propose the analogous Lorentz-violating action for a scalar field,
\begin{equation}
S=-\frac{1}{2}\int \ud^4x \ \left[\left(\eta^{ab}+c^{ab}\right)\nabla_a\phi\nabla_b\phi+m^2\phi^2\right],
    \label{scalarLagrangian}
\end{equation}
where $c^{ab}$ is the corresponding SME coefficient and $m$ is the field's mass. Also, it is worth introducing the Lagrangian density, $\mathcal L$, which satisfies $S=\int\ud^4 x\ \mathcal L$. Observe that there are two non-dynamical objects in the action: the background Minkowski metric, $\eta^{ab}$, which is Lorentz invariant, and $c^{ab}$. Naively, one could argue that the theory has only one non-dynamical effective metric, $\tilde{\eta}^{ab}\equiv\eta^{ab}+c^{ab}$. However, this view neglects that the volume element in the action is associated with $\eta_{ab}$.

We consider a particular $c^{ab}$ whose analogue in the SME literature is known as an isotropic $c$ coefficient. The components of this SME coefficient, in a global set of Minkowskian coordinates $(t,x,y,z)$, can be written as
\begin{equation}
c^{\mu\nu}=\mathrm{diag}(C,0,0,0),
\label{particular}
\end{equation}
where $C$ is a constant. We assume $C<1$ so that the equations of motion are hyperbolic. Clearly, $c^{ab}$ does not share the symmetries of $\eta_{ab}$. Specifically, its Lie derivative along a boost Killing field of $\eta_{ab}$ is non-zero. As a consequence, the theory is only invariant under a very restricted set of diffeomorphisms: $\eta_{ab}$ reduces this symmetry to rigid (global) spacetime translations and rotations (including boosts), and the chosen $c^{ab}$ precludes the boosts from being a symmetry of the theory.

From a pair of complexified solutions to the equations of motion, $\phi$ and $\psi$, we define the Klein-Gordon product in terms of the Lagrangian density as
\begin{equation}
    (\phi,\psi)=-\ui\int \ud\Sigma\left(\frac{\partial \mathcal L}{\partial(\partial_t\bar\phi)}\psi-\frac{\partial \mathcal L}{\partial(\partial_t\psi)}\bar\phi\right),
    \label{defkgprod}
\end{equation}
where integration is over any Cauchy surface with the appropriate volume element $\ud\Sigma$. 
Nevertheless, for $c^{ab}$ as in Eq.~\eqref{particular}, the result is independent of such a surface. To see this, note that Eq.~\eqref{defkgprod} for the theory at hand can be cast in the manifestly covariant form
\begin{equation}
(\phi,\psi)=-\ui\int\ud\Sigma_a\ \mathcal J^a(\phi,\psi),
\end{equation}
where 
\begin{equation}
\mathcal J^a(\phi,\psi)=\tilde\eta^{ab}(\psi\nabla_b\bar\phi-\bar\phi\nabla_b\psi),
\end{equation}
and $\ud\Sigma_a$ is the surface's volume form times the future-directed unit normal. Using Eq.~\eqref{particular}, we can show that $\nabla_a\mathcal J^a(\phi,\psi)=0$. Hence, an application of Gauss' theorem reveals that $(\phi,\psi)$ is independent of the Cauchy surface used for integration.

In the privileged coordinates, the Klein-Gordon product becomes
\begin{equation}
    (\phi,\psi)=-\ui(1-C)\int \ud^3x\  (\psi\partial_t\bar\phi-\bar\phi\partial_t\psi),
    \label{kg}
\end{equation}
where we integrate over a surface of constant $t$.
We utilise the modes\footnote{It is often useful to formally study fields with periodic spatial dimensions to have discrete momenta. However, this is inconvenient here since the boundary conditions themselves select a `preferred' frame.}
\begin{equation}
    \phi_{\vec k}=[(2\pi)^32\omega(1-C)]^{-1/2}\exp\left[{\ui\left(\vec k\cdot \vec x-\omega t\right)}\right],
    \label{privmodes}
\end{equation}
with
\begin{equation}
\omega \equiv \sqrt{\frac{m^2+\vec k^2}{1-C}},
    \label{privdisp}
\end{equation}
and where the dot represents the standard $3$-dimensional scalar product. These modes, together with their complex conjugates, comprise a complete set of solutions to the field equations and are normalised so that $(\phi_{\vec k_1},\phi_{\vec k_2})=\delta^{(3)}(\vec k_1-\vec k_2)$.

The modes~\eqref{privmodes} are adapted to the privileged coordinates in the sense that they are eigenfunctions of the Lie derivative along $\partial_t$, $\mathsterling_{\partial_t}$, with eigenvalues $-\ui\omega$ (recall that $\omega>0$). These modes also happen to be eigenfunctions of $\ui\mathsterling_{\partial_{t^\prime}}$, but the positivity of their eigenvalues is no longer guaranteed. Likewise, we can adopt a set of solutions with positive eigenvalues with respect to $\partial_{t^\prime}$; these are
\begin{equation}
\varphi_{\vec k^\prime}=\mathcal N_{\vec k^\prime}\exp\left[\ui\left(\vec k^\prime\cdot\vec x^\prime-\omega^\prime t^\prime\right)\right],
    \label{unpmodes}
\end{equation}
where $\mathcal N_{\vec k^\prime}$ are normalisation constants. The dispersion relation now solves
\begin{equation}
    (\omega^\prime)^2-(\vec k^\prime)^2-m^2+C\left[-\gamma^2(\omega^\prime)^2-2v\gamma k_z^\prime\omega^\prime-\gamma^2v^2(k_z^\prime)^2\right]=0,
    \label{undisp}
\end{equation}
where $\gamma\equiv 1/\sqrt{1-v^2}$ and we choose the boost direction to be $+z$, without loss of generality.

The Klein-Gordon product~\eqref{defkgprod} can be also written in terms of the primed coordinates,
\begin{equation}
    \left(\phi,\psi\right)^\prime=-\ui\int\ud^3 x^\prime\left\{\psi\left[(1-C-v^2)\gamma^2\partial_{t^\prime}\bar\phi+\gamma^2vC\partial_{z^\prime}\bar\phi\right]-\bar\phi\left[(1-C-v^2)\gamma^2\partial_{t^\prime}\psi+\gamma^2vC\partial_{z^\prime}\psi\right]\right\},
    \label{unkg}
\end{equation}
where integration is on a surface of constant $t^\prime$. From Eq.~\eqref{unpmodes}, we note that
\begin{equation}
\left(\varphi_{\vec k_1^\prime},\varphi_{\vec k_2^\prime}\right)^\prime=2\overline{N_{\vec k_1^\prime}}\mathcal N_{\vec k_2^\prime}(2\pi)^3
\left[(1-C-v^2)\gamma^2\omega_1^\prime-C\gamma^2vk_{z1}^\prime\right]\delta^{(3)}\left(\vec k_1^\prime-\vec k_2^\prime\right),
    \label{unpnorm}
\end{equation}
where $\omega_1^\prime$ satisfies the dispersion relation \eqref{undisp} with $\vec k_1^\prime$ instead of $\vec k^\prime$. From Eq.~\eqref{unpnorm}, we can read off the normalisation constants $\mathcal N_{\vec k^\prime}$. These are all the elements we need to perform a canonical quantisation of the Lorentz-violating scalar field and compare different inertial vacua, which is the main goal of the next section.

\section{Inequivalence of inertial vacua}
\label{Sec:Vacua}

In standard quantum field theory, the existence of a Lorentz-invariant vacuum is of fundamental importance, so much so that the existence of such a \emph{unique} Poincar\'e-invariant state was adopted as an \emph{axiom} in the axiomatic formulation of quantum field theory in Minkowski spacetime~\cite{Streater64}. Indeed, since in field theory the (Weyl) algebra of observables admits unitarily inequivalent representations (due to the inapplicability of the Stone-von Neumann theorem), the Poincar\'e symmetry serves as a guiding principle in selecting a distinguished one, i.e., the one in which observables are operators acting on the Fock space with the distinguished Minkowski vacuum. This property, as we shall see, is lost in Lorentz-violating theories.

We begin this section by noting that, since both sets of modes, $\{\phi_{\vec k},\bar\phi_{\vec k}\}$ and $\{\varphi_{\vec k^\prime},\bar\varphi_{\vec k^\prime}\}$ are complete, we may write
\begin{equation}
\phi_{\vec k}=\int \ud^3 k^\prime\left(\alpha_{\vec k\vec k^\prime}\varphi_{\vec k^\prime}+\beta_{\vec k\vec k^\prime}\bar\varphi_{\vec k^\prime}\right).
    \label{bogo}
\end{equation}
Let $\boldsymbol\Phi$ denote the field operator, then
\begin{equation}
    \boldsymbol\Phi=\int\ud^3  k \left(\phi_{\vec k}a_{\vec k}+ \mathrm{h.c.}\right)=\int\ud^3 k^\prime\left(\varphi_{\vec k^\prime}a^\prime_{\vec k^\prime}+\mathrm{h.c.}\right),
    \label{modeexp}
\end{equation}
where $\mathrm{h.c.}$ denotes the Hermitian conjugate. From Eq.~\eqref{modeexp} we can find the corresponding Bogoliubov coefficients~\cite{Bire82}
\begin{equation}
\alpha_{\vec k\vec k^\prime}=\left(\varphi_{\vec k^\prime},\phi_{\vec k}\right)^\prime,\qquad\beta_{\vec k\vec k^\prime}=-\left(\bar\varphi_{\vec k^\prime},\phi_{\vec k}\right)^\prime.
    \label{alphabeta}
\end{equation}
In addition, we can explicitly compute
\begin{multline}
\beta_{\vec k\vec k^\prime}=\frac{\ue^{-\ui(\omega^\prime-\gamma(\omega+v k_z))t^\prime}\delta(k_x-k_x^\prime)\delta(k_y-k_y^\prime)\delta(\gamma(k_z-\omega v)-k_z^\prime)}{2(2\pi)^3\sqrt{\omega(1-C)[(1-C-v^2)\gamma^2\omega^\prime-C\gamma^2v k_z^\prime]}}\\
\times\left[(1-C-v^2)\gamma^2(\omega^\prime+\gamma(k_zv-\omega))+C\gamma^2v(\gamma(k_z-\omega v)-k_z^\prime)\right].
    \label{beta}
\end{multline}
Note that these coefficients are only zero for 
\begin{equation}
k_z=v\sqrt{\frac{m^2+k_x^2+k_z^2}{1-C-v^2}},
\end{equation}
(as reflected in Fig.~\ref{fig:my_label} below). What is more, we can easily check that, if $C=0$ or $v=0$, then $\beta_{\vec k\vec k^\prime}=0$. Nevertheless, in general, $\beta_{\vec k\vec k^\prime}\neq 0$. Assuming $\beta_{\vec k\vec k^\prime}\neq0$ and denoting $|0\rangle$ the vacuum associated to $a_{\vec k}^\prime$, that is, the state such that $a_{\vec k}|0\rangle=0\ \forall\vec k$, we can readily verify that $a^\prime_{\vec k}|0\rangle\neq0$. Since the two relevant sets of modes are associated with different inertial frames, this result already shows that the vacuum is not Lorentz invariant in the theory we are considering.

Moreover, Eqs.~\eqref{bogo} and \eqref{alphabeta} imply 
\begin{equation}
a_{\vec k}^\prime=\int\ud^3 k^\prime\ \left(\alpha_{\vec k\vec k^\prime}a_{\vec k^\prime}+\overline{\beta_{\vec k\vec k^\prime}}a^\dagger_{\vec k^\prime}\right).
    \label{annihilation}
\end{equation}
We can use this last relation to show that the number operator, $N_{\vec k}^\prime=(a^\prime_{\vec k})^\dagger a^\prime_{\vec k}$, satisfies
\begin{equation}
\langle 0|N_{\vec k}^\prime|0\rangle=\int\ud^3 k^\prime \left|\beta_{\vec k\vec k^\prime}\right|^2,
    \label{number}
\end{equation}
where, in the second step, we have applied Eq.~\eqref{annihilation} and the commutation relation
\begin{equation}
    \left[a_{\vec k^\prime_1}, a^\dagger_{\vec{k_2}^\prime}\right]=\left(\phi_{\vec k_1^\prime},\phi_{\vec k_2^\prime}\right)\mathbf 1.
\label{commutation}
\end{equation}

Again, whenever the coefficients $\beta_{\vec k\vec{k^\prime}}$ are non-zero, the preferred vacuum $|0\rangle$ has a non-zero number of particles in the modes adapted to the unprivileged frame.
From Eqs.~\eqref{beta} and~\eqref{number}, we get the formal expression
\begin{equation}
   \langle 0|N^\prime_{\vec k}|0\rangle=\mathcal K(\vec k)\delta^{(3)}(\vec{0}),
   \label{number1h}
\end{equation}
where
\begin{equation}
\mathcal K(\vec k)=\frac{\gamma^2[(1-C-v^2)(\omega^\prime-\gamma(\omega-k_z v))-Cv(k_z^\prime-\gamma(k_z-\omega v))]^2}{\omega(1-C)[(1-C-v^2)\omega^\prime-Cvk_z^\prime]},
    \label{number2}
\end{equation}
and the presence of the Dirac delta implies that $ \langle 0|N^\prime_{\vec k}|0\rangle$ must be interpreted distributionally as follows: The $\beta_{\vec k\vec k^\prime}$ give rise to $\beta_{\vec k}(f)=\int \ud^3 k^\prime \beta_{\vec k\vec k^\prime} f(\vec k^\prime)$, where $f$ is a test function specifying a momentum profile. This implies that the number operator must be defined as a bi-distribution such that $\langle0|N^\prime_{\vec k}(f_1,f_2)|0\rangle=\beta_{\vec k}(f_1)\overline{\beta_{\vec k}(f_2)}$. Setting $f_1=f_2=f$, we get $\langle0|N^\prime_{\vec k}(f,f)|0\rangle =\mathcal K(\vec k)|f(k)|^2$.

Equation~\eqref{number2} is written with a small abuse of notation as $k_z^\prime$ is no longer free but rather obtained from $k_z$ by the corresponding Lorentz transformation, and, as before, $\omega$ and $\omega^\prime$ are calculated using their corresponding dispersion relations.

Notice that $\mathcal K(\vec k)=0$ for $C=0$ or $v=0$. On the other hand, $\mathcal K(\vec k)$
diverges when $C<0$ and the phase velocity in the `primed frame', $\omega^\prime/k_z^\prime$ reaches $Cv/(1-C-v^2)$. These effects are closely related to the vacuum \v{C}erenkov radiation, which has been studied in the context of the SME~\cite{VacCerenkov1, VacCerenkov2}. Another noteworthy feature of Eq.~\eqref{number2} is that the expectation value of the number operator approaches constant values for large $|k_z|$. Figure~\ref{fig:my_label} shows a plot of $\mathcal K(\vec k)$ for the particular case $k_x=k_y=0$.

\begin{figure}[t]
    \centering
    \includegraphics[width=.65\textwidth]{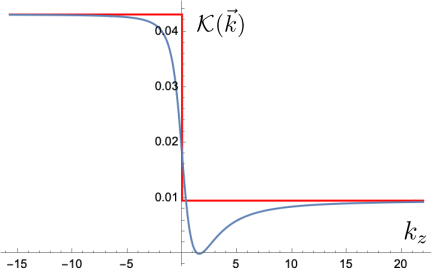}
    \caption{$\mathcal K(\vec k)$ as a function of $k_z$ for $m=1$ (blue curve) and $m=0$ (red curve), and with $k_x=k_y=0$. We used the parameters $C=0.1$ and $v=0.8$.
    Note that $\mathcal K(\vec k)$ approaches non-zero constants as $k_z\to\pm\infty$. }
    \label{fig:my_label}
\end{figure}

Furthermore, as done in the build-up of Theorem 4.4.1 of Ref.~\cite{WaldQFT}, we assume that there is a linear transformation linking the two inertial vacua. It turns out that, for this transformation to exist as a \emph{bona fide} unitary operator, we must have
\begin{equation}
    \displaystyle\mathrm{Tr}\int\ud^3 k_3\ \beta_{\vec k_1\vec k_3}\overline{\beta_{\vec k_3 \vec k_2}}<\infty.
\end{equation}
However, Eq.~\eqref{number} says otherwise, as can be seen by inspecting the asymptotic behaviour of $\beta_{\vec k_i\vec k_j}$ for large values of $|k_z|$. This shows that the Fock spaces constructed from the one-particle Hilbert space adapted to modes~\eqref{unpmodes} and~\eqref{privmodes} are not unitarily equivalent. In turn, this observation is related to our dealing with fields, which have infinitely many degrees of freedom; in theories with finite values of $\vec k$, the celebrated Stone-von Neumann theorem guarantees the existence of the unitary transformation.

We now turn to study the interaction of the Lorentz-violating scalar field with inertial Unruh-De Witt detectors.

\section{Inertial Unruh-De Witt detectors}
\label{Sec:Detector}

If one wishes to extract observable consequences of a theory, one must consider interactions. The simplest model that captures the relevant physics is an Unruh-De Witt detector~\cite{Bire82}: a pointlike two-level quantum system that follows a spacetime worldline as it interacts with the quantum field. The energy eigenvectors of the detector form a basis for its two-level Hilbert space. We denote them by $|E_0\rangle $ and $|E_1 \rangle$ with energies $E_0$ and $E_1>E_0$, respectively.

Assuming that the detector and the field are constantly coupled for an arbitrarily large proper detector time, we can write the interaction Hamiltonian as
\begin{equation}
H_I=\lambda\mu(\tau)\otimes\phi(x(\tau)),
    \label{UDWscalar}
\end{equation}
where $\mu$ denotes an internal detector degree of freedom, usually known as its monopole moment, $\lambda$ is a coupling constant, and $\phi(x(\tau))$ denotes pullback of the filed to the detector's worldline, parameterised by its proper time $\tau$. Since the detector moves inertially, its worldline can be described as $(t,x,y,z)=(\gamma\tau,0,0,vt)$.

To first order in $\lambda$, the transition amplitude from the detector's ground state $|E_0\rangle$ to the excited state $|E_1\rangle$ is~\cite{Unruh76}
\[ A^{(1)}\equiv -\ui\lambda\langle E_1|\langle\boldsymbol\Psi|\int_{-\infty}^\infty\ud\tau\ \mu(\tau)\phi(x(\tau))\ |0\rangle|E_0\rangle,\]
where $|\boldsymbol\Psi\rangle$ is the field state and $|0\rangle$ denotes the `privileged' vacuum, as before. This last equation factorises as
\begin{equation}
  A^{(1)}=  -\ui\lambda\langle E_1|\mu(0)|E_0\rangle\int\ud\tau\ \ue^{\ui(E_1-E_0)\tau}\langle\boldsymbol\Psi|\phi(x(\tau))|0\rangle,
   \label{Tamplitude}
\end{equation}
so the only non-zero matrix elements are
\[M_k\equiv\langle 0|a_k\phi(x(\tau))|0\rangle=[(2\pi)^32\omega(1-C)]^{-1/2}\ue^{\ui(\omega t-\vec k\cdot\vec x)},\]
where, in the second step, we used the expansion in modes given in Eq.~\eqref{privmodes}.

We can straightforwardly verify that the other terms in the Dyson series~\cite{Reed75} share the form
\begin{multline}
 A^{(n)}\equiv   \frac{(-\ui\lambda)^n}{n!}\langle E_1|\langle\boldsymbol\Psi|\int\ud\tau_1\int\ud\tau_2\ \cdots\ \int\ud\tau_n\ \theta(\tau_{n}-\tau_{n-1})\cdots\theta(\tau_2-\tau_1)\\
 \times H_I(x(\tau_1))\cdots H_I(x(\tau_n))|0\rangle|E_0\rangle.
    \label{odern}
\end{multline}

Thus, to order $n>1$, all the matrix elements for which $\langle\boldsymbol\Psi|$ is not the product of $n$ creation operators, vanish identically. The nontrivial matrix elements are
\begin{multline}
  A^{(n)}=      \frac{(-\ui\lambda)^n}{n!}\langle E_1|\mu(0)^n|E_0\rangle
        \int_{-\infty}^\infty\ud\xi\ \ue^{\ui[(E_1-E_0)+\frac{\gamma}{n}\sum_{j=1}^n(\omega_j-\vec k_j\cdot\vec v)]\xi}\times\\
      \int_0^\infty\prod_{i=1}^{n-1}\ud\eta_i\ \frac{(-1)^{n+1}}{n}\ \sum_{\sigma,i,l}\mathrm{sgn}(\sigma)\ \sigma\left(\frac{\exp\left\{-\ui\gamma[\sum_{r=1}^{n-1}k_lx(P^{-1}_{i,r+1}\eta_r)]\right\}}{\sqrt{(2\pi)^32\omega_i(1-C)}}\right),
        \label{interm}
\end{multline}
on which we introduce the variables $\xi\equiv \sum_{i=1}^n\tau_i$ and $\eta_i\equiv \tau_i-\tau_{i+1}$ for $i=1,2,\ldots,n-1$. Here, $\sigma$ denotes a permutation of $i$ and $l$, $\mathrm{sgn}(\sigma)$ is $1$ (resp. $-1$) for even (resp. odd) permutations, and $P$ denotes the matrix implementing the transformation form $(\xi,\eta_i)$ to $\tau_i$, which is given by
\[P=\begin{bmatrix}
1 & 1 & 1 & \cdots & 1\\
1 & -1 & 0 & \cdots & 0\\
0 & 1 & -1 & \cdots & 0\\
\vdots & \vdots & \vdots & \ddots& \vdots\\
0 & 0 & 0 & \cdots & -1
\end{bmatrix}.\]
We also use that $\det P=(-1)^{n+1}n$ and $P_{i1}^{-1}=1/n$ for each $i$. 

When we integrate over $\xi$, we can see that, for all $n$,
\begin{equation}
A^{(n)}\propto \delta\left(E_1-E_0+\frac{\gamma}{n}\sum_{j=1}^n(\omega_j-\vec k_j\cdot\vec v)\right).\label{delta}
\end{equation}
Given that $E_1>E_0$, if
\begin{equation}
v<\frac{\omega_j}{k_j} \qquad\forall j,
\label{bound}
\end{equation}
then, the argument of the delta function in Eq.~\eqref{delta} is strictly positive, and, consequently, the inertial detector remains silent. Recall that this result is valid for all orders in the Dyson series. Physically, condition~\eqref{bound} means that the detector moves slower than the modes' phase velocity\footnote{Since we have linear dispersion relation, the phase and group velocities coincide.}. A particular realisation of condition~\eqref{bound} is when the phase velocities are bounded by the speed of light, in which case Eq.~\eqref{bound} holds for all $v<1$. In our example, the necessary and sufficient condition for this realisation is $0\leq C<1$\footnote{Indeed, let $\alpha\equiv 1/\sqrt{1-C}$. The condition we wish to satisfy is $\alpha\sqrt{\vec k^2+m^2}\geq|\vec k|\ \forall \vec k$. Since the left-hand side is a convex function which is larger than $|\vec k|$ for $|\vec k|=0$, we need the equation $\alpha\sqrt{\vec k^2+m^2}=|\vec k|$ to have no real solutions in this range. Writing the discriminant of the square of the equation, we see that this happens for $\alpha\geq 1$.}.

An alternative procedure to look for observable consequences of Lorentz violation consists in violating the bound~\eqref{bound}. A direct calculation of the Wightman function written in terms of the privileged coordinates in the massless case reveals a divergence when $v=1/\sqrt{1-C}$. In the detector model, such divergence can influence the rate of spontaneous de-excitation~\cite{Benito17}, but this is a regime where other effects, like vacuum \v{C}erenkov radiation~\cite{VacCerenkov1, VacCerenkov2}, are expected to occur.

Of course, there are limitations when using Unruh-De Witt detectors. Nevertheless, they function as a technically simple tool to identify interesting cases. For example, if the Lorentz-violating sector responds to an effective metric $\tilde\eta_{ab}$ that is not flat, the transition amplitudes~\eqref{Tamplitude} could be non-zero. Likewise, non-inertial detectors could motivate new phenomenological tests of Lorentz violation; these tests require that we first develop the experimental capacity to measure the Unruh effect~\cite{Good:2020hav, Gooding:2020scc}.

This section shows that, despite an `unprivileged' inertial detector describing the vacuum $|0\rangle$ as a sea of particles following the distribution~\eqref{number2}, it does not interact with them, except for detectors that move as fast as the field modes. We turn now to analyse the singular structure of the two-point function in the theory at hand.

\section{The Hadamard property in theories with Lorentz violation}\label{sec:Hadamard}

A property of central importance for free and perturbatively-interacting quantum field theories is the Hadamard condition, which characterises the distributional singularities of the two-point function for physically reasonable states~\cite{Fewster:2013lqa}. In this section, we examine how the Hadamard condition becomes altered for the Lorentz-violating theory under consideration. The Hadamard property is typically formulated in terms of the Wightman function, which, for a state vector $\Psi$ in a concrete representation, is $\omega_2(x,x') = \langle \Psi | {\bf \Phi}(x) {\bf \Phi}(x') | \Psi \rangle$.

For the theory defined by the action~\eqref{scalarLagrangian}, the Wightman function is a bi-solution of the field equation
\begin{equation}
P \otimes 1 \omega_2 = 1 \otimes P \omega_2 = 0,
\label{P-op-bi}
\end{equation}
where the differential operator, $P$, is defined as
\begin{equation}
P = \left(\eta^{ab}+c^{ab}\right)\nabla_a\nabla_b-m^2 = \tilde \eta_{ab} \nabla_a \nabla_b - m^2.    
\label{P-op}
\end{equation}

Direct inspection of the characteristic set of the differential operator $P$ indicates that the standard Hadamard property does not hold in Lorentz-violating theories. Indeed,
the principal symbol of $P$ is given by $p(x,k) = \tilde \eta^{ab}(x) k_a k_b$ and hence the characteristic set,
\begin{equation}
\mathcal{N} = {\rm Char} P = \{ (x,k) \in \mathbb{R}^4 \times \mathbb{R}^{4 *}\setminus{0}: \eta^{ab}(x) k_a k_b = 0  \},
\end{equation}
lies along non-vanishing null co-vectors with respect to the effective metric $\tilde \eta_{ab}$, and not along the standard (non-vanishing) null co-vectors of Minkowski spacetime.

The above observations suggest that it is convenient to study the Hadamard property from the microlocal viewpoint introduced by Radzikowski~\cite{Radzikowski}, which might be adapted more straightforwardly to the Lorentz-violating case. Also, it appears to us that this has the advantage that the result can be directly generalised to a large class of Lorentz-violating theories. By a standard result in microlocal analysis~\cite{Hormander}, any distribution defined on smooth functions of compact support, $\varphi: C^\infty_0 (\mathbb{R}^4) \to \mathbb{R}$, that satisfies $P \varphi = 0$, must have a wavefront set\footnote{Let $\varphi$ be a distribution. We call $(x, k)$ with $k \neq 0$ a \emph{regular direction} of $\varphi$ if one can find $u \in C_0^\infty(\mathbb{R}^4)$ with support including $x$, such that the Fourier transform of the pointwise product $\varphi u$ decays rapidly in a conic neighbourhood of $k$. The wavefront set of $\varphi$, denoted by ${\rm WF}(\varphi)$, is defined as the set of pairs $(x,k)$ with $k \neq 0$, such that $(x,k)$ are not regular directions of $\varphi$.} satisfying 
\begin{align}
    {\rm WF} (\varphi) \subset \mathcal{N}.
    \label{WFset1}
\end{align}
In fact, this analysis is valid for any second-order hyperbolic differential operator in Lorentz-violating theories whose principal symbol is given in terms of an effective Lorentz-violating metric, analogous to $\tilde \eta$ in Eq.~\eqref{P-op}. Thus, we can continue our discussion in full generality.
 
From Eq.~\eqref{P-op}, it can be verified that any distributional singularities of a distributional solution of the Lorentz-violating Klein-Gordon equation propagate along the `effective' null cones of $\tilde \eta_{ab}$, where, recall that we work under the assumption that $c_{ab}$ does not alter the Lorentzian character of $\tilde \eta_{ab}$. This result can be extended straightforwardly to distributional bi-solutions such as $\omega_2$. If $\varpi$ is a distributional bi-solution, one is now led to a wavefront set condition satisfying
\begin{align}
{\rm WF}(\varpi) \subset {\cal N}_0 \times {\cal N}_0, 
\end{align}
where ${\cal N}_0$ is equal to ${\cal N}$, but including zero co-vectors, i.e., elements of the form $(x,0)$. 

With this in mind, a direct generalisation of the Lorentz-invariant case suggests that the correct microlocal spectrum condition for the (positive frequency) Wightman function, $\omega_2$, should further satisfy
\begin{equation}
{\rm WF}(\omega_2) \subset \cal N^+ \times \cal N^-,
\end{equation}
where $\cal N^+$ is the set of co-vectors in $\cal N$ with future-pointing momenta and $\cal N^-$ is the set of co-vectors in $\cal N$ with past-pointing momenta~\cite{Radzikowski}.

So far, all the results for characterising the wavefront set of $\omega_2$ are as in the Lorentz invariant case, but replacing the Minkowski metric by the effective metric $\tilde{\eta}_{ab}$. We can therefore directly write down the microlocal spectrum condition~\cite{Radzikowski} for Lorentz-violating theories
\begin{multline}
{\rm WF}(\omega_2)  = \{ ((x_1, k_1),(x_2 k_2)) \in (\mathbb{R}^4 \times \mathbb{R}^{4 *}\setminus{0}) 
 \times (\mathbb{R}^4 \times \mathbb{R}^{4 *}\setminus{0}):\\
 (x_1,k_1) \sim (x_2, -k_2),
 \text{ and } k_1 \text{ is future pointing}\},
\label{MLSC}
\end{multline}
where the notation here is that $(x,k) \sim (y,p)$ if there is a curve $\gamma$ whose image contains the points $x$ and $y$, such that the tangent co-vector field to $\gamma$ satisfies $\tilde \eta^{ab} v_a v_b =0$ at every point along the curve, and that $k$ and $p$ are parallel transports along $\gamma$. Note that here we use the effective metric and not the Minkowski metric.

In support of Eq.~\eqref{MLSC}, we can write in closed form the two-point function for the theory studied in the previous sections. We have
\begin{align}
\omega_2(x,x') = \frac{m}{4 \pi^2} \frac{K_1\left(m \sqrt{-(1-C)^{-1}(t-t' - i \epsilon )^2 + (\vec x - \vec x') \cdot (\vec x - \vec x') + \epsilon^2 } \right)}{\sqrt{-(1-C)^{-1}(t-t' - i \epsilon )^2 + (\vec x - \vec x') \cdot (\vec x - \vec x') + \epsilon^2 }}.
\end{align}
The factor $m/(4\pi^2)$ is chosen so that in the $C = 0$ case one recovers the Minkowski state Wightman function. If we define $\tilde\sigma = [-(1-C)^{-1}(t-t' )^2 + (\vec x - \vec x') \cdot (\vec x - \vec x')]/2$, we can see straightforwardly that the modified Hadamard structure takes the form
\begin{align}
\omega_2(x,x') = \frac{1}{8 \pi^2 \tilde \sigma }+\frac{m^2}{16 \pi^2}  \left[ \log \left(m^2 \tilde \sigma/2 \right)+2 \gamma -1\right] + O\left[\tilde \sigma \log (m \tilde \sigma) \right],
\label{ModHad}
\end{align}
where $\gamma$ is Euler's gamma, and it is readily seen that the singular structure is, as we have claimed above, associated with the causal structure of the effective metric $\tilde \eta_{ab}$, rather than to the Minkowski metric $\eta_{ab}$. The form of expansion~\eqref{ModHad} is particularly appealing because, for $C = 0$, it coincides exactly with the Hadamard expansion of the Poincar\'e-invariant Minkowski vacuum. 

From the expansion~\eqref{ModHad}, it is clear how to renormalise non-linear observables in this theory via point-splitting and Hadamard subtraction. For example, the renormalised expectation value of ${\bf \Phi}^2$ is
\begin{align}
\langle 0 | {\bf \Phi}^2 | 0 \rangle = \frac{m^2(2 \gamma -1)}{16 \pi^2} +  \alpha m^2,
\end{align}
where $\alpha$ is a dimensionless renormalisation ambiguity, which allows one to fix $\langle 0 | {\bf \Phi}^2| 0 \rangle = 0$. By a similar argument, we can set the expectation value of the stress-energy tensor in the inertial vacuum $|0\rangle$ to zero, which is an axiomatic requirement imposed on the Poincar\'e-invariant Minkowski vacuum~\cite{WaldQFT}.

What is more, it can be inferred that generalisations of the Hadamard singular structure to Lorentz-violating theories, described by effective metrics that differ from the spacetime metric, will continue to have a familiar Hadamard expansion generalising the flat spacetime case that we have presented here, which should be in agreement with the microlocal analytic expectation characterising the Hadamard singularities.

The curved spacetime generalisation is interesting in several ways. For instance, in locally Lorentz-invariant curved spacetimes it is natural to ask whether important effects occur, for instance, upon approaching Cauchy horizons\footnote{When considering regions of flat spacetimes, one can already discuss phenomena associated to Cauchy horizons, as occurs when considering a Rindler wedge as a spacetime~\cite{Juarez-Aubry:2015dla}.}  or at the crossing of event horizons. In the former case, it is possible that the Hadamard condition can break down (see Refs.~\cite{Juarez-Aubry:2015dla, Juarez-Aubry:2021tae} and references therein). In the latter case, one needs to find a reasonable choice of states that leads to consistent physics~\cite{Juarez-Aubry:2014jba, Juarez-Aubry:2018ofz}. However, in a large class of Lorentz-violating theories, one can find geometrically interesting regions, like universal horizons~\cite{Berglund:2012bu} or Cauchy horizons associated with the effective metrics of the theory. In view of our characterisation of the modified Hadamard property in Lorentz-violating theories, we consider that the exploration of quantum phenomena, in these regions, is worthy of further exploration.

As a final comment, the renormalisation ambiguities in the construction of non-linear observables in quantum field theory in curved spacetimes have been classified in terms of a series of axioms relying crucially on the geometry and scaling properties of the corresponding spacetime~\cite{Hollands:2001nf, Hollands:2001fb}. It appears to us that in Lorentz-violating theories, which include additional (dynamical or non-dynamical) fields, the classification of ambiguities ought to be expanded. In particular, this classification should include, in addition to geometrical quantities, quantities constructed with the additional fields of the theory. This deserves to be further investigated.

\section{Discussion}\label{sec:conc}

We analysed the quantisation of a simple Lorentz-violating scalar theory. We found that the vacuum, as perceived by an inertial observer, is populated from the viewpoint of a different inertial frame, refuting the presumption that the vacuum is always unique and Lorentz invariant. This result could have had profound empirical consequences had it entailed physical effects on observers moving in different inertial frames. To this end, we examined the response of an inertial Unruh-DeWitt detector and showed, however, that it is insensitive ---to all orders in perturbation theory--- to the excitations of this scalar field. Following the maxim `a particle is what a particle detector detects', the vacuum appears empty in all inertial frames, after all. So, the answers to the questions in the title are negative, at least for the model we studied.

In addition, we discussed how the Hadamard property needs to be adapted in any theory whose effective light cone does not coincide with that of the Minkowski metric, finding {\it a priori} no obstructions to treating Lorentz-violating fields in interacting quantum theory or for the construction of renormalised non-linear observables. Studying interacting Lorentz-violating fields and constructing non-linear observables are interesting open questions that can be examined in the future as a test of the internal mathematical consistency of Lorentz-violating theories.

Our study could be generalised to other maximally symmetric spacetimes where there are preferred vacua. More interestingly, we believe that the presented goals can be analysed in more realistic field theories without finding major obstacles. Notice that more general SME coefficients (e.g., spacetime-dependent coefficients) could have detectable signals associated with the inequivalence of the inertial vacua, which could set the ground for promising experimental tests. In light of the lessons of this work, we urge the proponents of particular Lorentz-violating models to consider the corresponding interacting quantum field theory and verify their mathematical and empirical adequacy. It is our hope that the present work will also lead to stimulating new directions of research in `curved spacetimes' Lorentz-violating theories including Ho\v{r}ava-Lifshitz or Einstein-Aether gravity.

\section*{Acknowledgments}
The authors wish to thank Lukas Nellen for his helpful ideas and Jorma Louko for pointing out an issue in an earlier version of this article. This research was partially supported by UNAM-DGAPA-PAPIIT Grant No. IG100120, CONACyT FORDECYT-PRONACES Grant No. 140630. BAC is funded by a UNAM-DGAPA Postdoctoral Fellowship, BAJ-A is funded by a CONACYT Postdoctoral Fellowship.



\bibliographystyle{unsrt}
\bibliography{apssamp.bib}

\end{document}